\documentclass[prl]{revtex4}
\begin{document}
\title{Testing models for structural relaxation of
viscous liquids by temperature jump experiments}
\author{N. B. Olsen, T. Christensen, and J. C. Dyre}
\address{Department of Mathematics and Physics 
(IMFUFA), Roskilde University, Postbox 260, DK-4000 
Roskilde, Denmark}
\begin{abstract}
Based on the Tool-Narayanaswami formalism for 
structural relaxation we propose an experiment which
independent of the relaxation time distribution
determines the structural relaxation rate by subtracting 
relaxations following two different temperature jumps. 
In particular, the experiment makes it possible to 
evaluate the relaxation rate immediately
following a temperature jump - thus allowing one to
distinguish the entropy model from others.
\end{abstract}
\maketitle
The glass transition takes place when the relaxation time of a 
viscous liquid upon cooling becomes comparable to the cooling 
time. What determines the relaxation time is still a matter of 
debate. In this note \cite{note} we discuss an experiment which 
makes it possible 
to distinguish between two fundamentally different 
types of models for the relaxation time. 

Before proceeding let us briefly mention three phenomenological 
models for the relaxation time. These models are often just discussed
in connection with linear experiments, i.e., experiments monitoring
the equilibrium liquid, but all three models have obvious extensions 
to deal with structural relaxation. 
Writing for the relaxation time $\tau$

\begin{equation}\label{1}
\tau\ =\ \tau_0\exp(W)\,,
\end{equation}
the three phenomenological models are: The entropy model 
($W\propto 1/TS_c$ where $S_c$ is the configurational 
entropy per molecule), the free 
volume model ($W\propto 1/v_f$ where $v_f$ is the free volume
per molecule), and the shoving model \cite{shoving} ($W\propto G_\infty/T$ 
where $G_\infty$ is the instantaneous shear modulus). 
The quantities $S_c$, $v_f$ and $G_\infty$ are all 
temperature dependent - this is how the models are supposed 
to explain the non-Arrhenius temperature dependence of $\tau$. 

The entropy model differs from the two others in 
one important respect: Suppose temperature is changed 
``instantaneously,'' i.e. , much faster than the liquid relaxation 
time. Clearly, the quantity $S_c$ cannot change instantaneously. In 
contrast, both $v_f$ and $G_\infty$ have ``glassy'' 
contributions to their temperature dependence and thus do change with 
temperature even on a very fast time scale. 
Consequently, a study of how the relaxation rate 
changes after an instantaneous temperature jump makes it 
possible to distinguish the entropy model from the two other models.
Such an experiment answers the question: {\it How large is the instantaneous
change of the relaxation rate?}

Structural relaxation is generally nonlinear and thus much less 
well established theoretically than linear response theory. 
Nevertheless there exists a simple and beautiful extension of 
linear response theory to structural relaxation, applicable as 
long as only reasonably small temperature jumps are involved. 
This is the Tool-Narayanaswami (TN) formalism. 
The idea is that linear response theory still applies, except 
that time $t$ should be replaced by ``reduced'' time $\tilde{t}$. The 
reduced time is the time measured in units of the relaxation time 
(which in general itself changes with time during the 
experiment). More precisely, the reduced time is defined by

\begin{equation}\label{2}
d\tilde{t}\ =\ \frac{dt}{\tau(W(t))}\,.
\end{equation}
Suppose $Y$ is a quantity which is measured in 
an experiment where the temperature varies. Writing 
$Y=Y_0+Y_r$ where $Y_0$ is the part of $Y$ that changes 
instantaneously with temperature and $Y_r$ is the ``relaxing'' part 
of $Y$, the TN formalism predicts that a function $\phi$ exists 
such that

\begin{equation}\label{3}
\Delta Y_r(\tilde{t})\ =\ 
\int_0^\infty d\tilde{t}'\ \phi(\tilde{t}')\ 
\Delta T(\tilde{t}-
\tilde{t}')\,.
\end{equation}
Here $\Delta Y_r$ is the change of $Y_r$ from some reference 
state in themal equilibrium at the beginning of the experiment 
(and similarly for $\Delta T$). 

We shall only consider temperature jump experiments, i.e., where 
the system is in thermal equilibrium until at $t=0$ the temperature 
changes suddenly. In such experiments, if the variable $X$ is 
defined by $X(t)=C \Delta Y(t)/\Delta T$ where $C$
is determined from the condition $X\rightarrow 1$ as $t\rightarrow\infty$, 
the TN formalism implies that a function $f$ exists such that

\begin{equation}\label{4}
X(t)\ =\ 
f(\tilde{t})\,.
\end{equation}
It is straightforward to express $f$ in terms of $\phi$
and the instantaneous jump of $X$. 
Note that $X$ includes the instantaneous contribution to 
$\Delta Y$, $\Delta Y_0$, and thus the function $f$ begins 
at a nonzero value for $\tilde{t}=0$.

Equation (\ref{4}) implies

\begin{equation}\label{6}
\frac{dX}{dt}\ =\ 
\frac{f'(\tilde{t})}{\tau(W(t))},
\end{equation}
which in turn via Eq.\ (\ref{1}) implies

\begin{equation}\label{7}
\ln\left(\frac{dX}{dt}\right)\ =\ 
\ln[f'(\tilde{t})/\tau_0]-W
\end{equation}

Consider now two temperature jump experiments. Each obey 
Eq.\ (\ref{7}) and consequently by subtraction at the same 
reduced time 

\begin{equation}\label{8}
\Delta\ln\left(\frac{dX}{dt}\right)\ =\ 
-\Delta W\,.
\end{equation}
Since the measured quantity $X$ is a function of reduced time (Eq.\ (\ref{4})), in 
practice the subtraction is performed by subtracting the two experiments {\it at the same 
value of $X$}. The purpose of the subtraction, of course, is to eliminate $f$.

Equation (\ref{8}) is valid for the subtraction of any
two temperature jump experiments. A particularly 
simple situation, however, arises when the first experiment is an 
``infinitesimal'' structural relaxation experiment, i.e., with a 
temperature step so small that $W$ does not change significantly 
during the entire experiment. In this case the left hand side of 
Eq.\ (\ref{8}) is essentially $W$ of the second (nonlinear) 
experiment. Extrapolating to zero time gives us the instantaneous
contribution to $W$.

Having established a method for the determination of
$W$, we return to the comparison of the entropy model to the
two others. What is the prediction of the entropy model for the
instantaneous contribution to $W$?
Suppose the nonlinear experiment is a jump from temperature $T_1$ 
to $T_2$. From an ``infinitesimal'' structural relaxation 
experiment around $T_1$ the equilibrium value of $W$ at 
$T_1$, $W_1$, is known. 
Plotting the left hand side of Eq.\ (\ref{8}) as a function of 
$X$ now makes it possible to investigate how large  
$W$'s instantaneous contribution is: According to the entropy
model, right after $t=0$ $W$ is 
equal to $W_1T_1/T_2$, because for 
this model only the $T$ in the expression for $W$ can change 
instantaneously. If a larger instantaneous contribution is
found, the entropy model in its extension to deal with 
structural relaxation is not viable.

To test the entropy model we recently performed 
an experiment along the lines presented above \cite{note}
[talk given by N. B. Olsen at the Fourth International Discussion
Meeting on Relaxations in Complex Systems, Crete, June 18.-23., 2001].
The experiment is an improved version of one described in 
\cite{shovingnonlin}. The quantity monitored during the structural 
relaxation, $Y$,
is the instantaneous shear modulus, which can be measured by a
resonance technique with 20 ppm
relative accuracy (5\% absolute accuracy). With the new reduced-size 
piezo-ceramic transducer, the time constant for the thermal
setup is now only 20 seconds. The experiment was done on the silicone 
oil DC704 with temperatures during the experiment in the range 207-208 K. 
The nonlinear temperature jump was 1.5 K, the
``infinitesimal'' jump was 0.15 K. The cryostate is able to keep
the temperature constant within 0.1 mK over months; the actual
experiment monitored three structural relaxations, each lasting
10 days. - The experiment
showed that there is an instantaneous contribution to $W$ which is
twice as large as that predicted by the entropy model. On the other
hand, the results are consistent with the shoving model.


\begin{references}
\bibitem{note} We here briefly discuss a few points which will be 
considered in more detail as part of a forthcoming publication.
\bibitem{shoving} J. C. Dyre, N. B. Olsen, and T. Christensen, Phys. Rev. B 
{\bf 53}, 2171 (1996); J. C. Dyre, J. Non-Cryst. Solids {\bf 235-237}, 142 (1998).
\bibitem{shovingnonlin} N. B. Olsen, J. C. Dyre, and T. Christensen,
Phys. Rev. Lett. {\bf 81}, 1031 (1998).
\end{references}
\end{document}